\documentstyle[aps,multicol,pre,epsfig]{revtex}
\begin{document}
\def\sech{\mathop{\rm sech}\nolimits}
\def\csch{\mathop{\rm csch}\nolimits}
\def\coth{\mathop{\rm coth}\nolimits}
\def\span{\mathop{\rm Span}\nolimits}
\def\sn{\mathop{\rm sn}\nolimits}
\def\cn{\mathop{\rm cn}\nolimits}
\def\dn{\mathop{\rm dn}\nolimits}
\def\sign{\mathop{\rm sign}\nolimits}

\setlength{\textwidth}{180mm} \setlength{\textheight}{240mm}
\setlength{\parskip}{2mm}

\begin{title}
{\bf Instabilities and Bifurcations of Nonlinear Impurity Modes}
\end{title}


\author{Panayotis G. Kevrekidis$^1$, Yuri S. Kivshar$^2$, and Alexander S. Kovalev$^3$}

\address{$^1$Department of Mathematics and Statistics, University of Massachusetts,
Amherst MA 01003-4515, USA \\
$^2$ Nonlinear Physics Group, Research School of Physical Sciences
and Engineering, Australian National University,\\ Canberra ACT
0200, Australia \\
$^3$ Institute for Low Temperature Physics and Engineering,
Kharkov 61103, Ukraine}

\maketitle

\begin{abstract}
We study the structure and stability of nonlinear impurity modes
in the discrete nonlinear Schr{\"o}dinger equation with a single
on-site nonlinear impurity emphasizing the effects of interplay
between discreteness, nonlinearity and disorder. We show how the
interaction of a nonlinear localized mode (a discrete soliton or
discrete breather) with a repulsive impurity generates a family of
stationary states near the impurity site, as well as examine both
theoretical and numerical criteria for the transition between
different localized states via a cascade of bifurcations.
\end{abstract}

\pacs{PACS numbers: 03.40.Kf; }

\begin{multicols}{2}

\narrowtext

\section{Introduction}

Nonlinear localized modes in discrete systems, also called
intrinsic localized modes or discrete breathers, are
self-localized states that exist due to the interplay between
lattice coupling and nonlinear effects. Spatially localized modes
of the discrete nonlinear Schr{\"o}dinger (DNLS) equation, known
as {\em discrete solitons}, have appeared in many diverse areas of
physics, such as biophysics, nonlinear optics, solid state physics
\cite{rev} and, more recently, in the studies of the Bose-Einstein
condensates in optical lattices \cite{Tro_Sme01} and
photonic-crystal waveguides and circuits \cite{PBGs}.

In application to nonlinear guided-wave optics, discrete solitons
have been suggested to exist in nonlinear waveguide arrays
\cite{Chr_Jos88}, and they have been observed a decade later in
AlGaAs waveguide structures \cite{Eis_et98} (see also a review
paper \cite{rev}). Several other issues, such as the
discreteness-induced soliton dynamics and soliton interaction,
have been investigated. Regarding applications, it was recently
shown that discrete solitons in two-dimensional networks of
nonlinear waveguide arrays can be used to realize various
functional operations, such as blocking, routing, logic functions,
time gating, etc. \cite{Chr_Eug01}.

On the other hand, localized impurities are known to play a
crucial role in numerous physical systems, not only introducing
interesting wave scattering phenomena \cite{marad}, but also
creating the possibility for the excitation of impurity modes,
which are spatially localized oscillatory states at the impurity
sites \cite{Lif}. The relevant phenomenology has been recognized
as important in a variety of physical settings ranging from defect
modes in superconductors \cite{andreev} to the dynamics of the
electron-phonon interactions \cite{tsironis} and from the
propagation of light in dielectric super-lattices with embedded
defect layers \cite{soukoulis} to defect modes in photonic
crystals \cite{Joann}.

It is crucially important to examine the interplay between
disorder and nonlinearity. For the continuum problems, this
interplay is known to lead to the existence of symmetric and
asymmetric impurity modes and their interesting stability
properties (see Refs. \cite{bogdan,sukh} and references therein).
Here we concentrate on the study of the discrete systems such as
the DNLS model. Some earlier results for the discrete lattices
focused on the scattering of localized modes by impurities in the
DNLS equation \cite{forinash,Kro_Kiv96}, as well as in the
Klein-Gordon model \cite{archilla}. The effect of the impurity on
localization properties of the nonlinear lattice and stability of
localized modes was studied, to the best of our knowledge,
only for lattices with anharmonic
coupling \cite{KFK}.

However, the earlier theoretical predictions \cite{Kro_Kiv96} and
recent experimental results \cite{Pes_et99,Mor_et02} for the
interaction of the discrete solitons with structural defects in
arrays of AlGaAs optical waveguides suggest that discreteness
should play a crucial role in the properties and stability of
nonlinear localized modes. In this paper, we study, in the
framework of the DNLS model, how the interaction of a discrete
localized mode with a repulsive nonlinear impurity leads to
different localized states, as well as examine, both analytically
and numerically, the stability criteria for the bifurcations
between different localized modes near the impurity site.

The existence of multiple localized states near the impurity site
can be understood by means of simple physics. Indeed, in the
continuum approximation, a two-hump localized mode centered at the
defect \cite{kos_kov} is known to be unstable with respect to an
exponential growth of antisymmetric linear perturbations
\cite{bogdan,sukh}. The above-mentioned instability leads to the
motion of the localized mode away from the defect \cite{sukh}, and
for small distances $\xi$ between the mode center and the defect,
the effective interaction energy can be presented in the form
$H_{\rm int} \simeq H_{0} - A {\xi}^2$,  where $A$ is defined by
the defect parameters. However, this situation becomes quite
different in discrete systems because of the so-called ``Peierls
relief'', an effective periodic potential due to the lattice
discreteness.  Such potential was first calculated for
small-amplitude breathers in a nonlinear elastic chain described
by the discrete nonlinear Klein-Gordon equation \cite{kov_bog}
applying an approach suggested for kinks of the Frenkel-Kontorova
model \cite{inden}. The potential has the following structure,
$H_{p} \simeq H_{1} - B \cos (2 \pi \xi / h)$, where $h$ is the
lattice spacing. Thus, a competition between these two potentials
defines a sequence of stable positions of a localized mode near
the defect. This qualitative picture is valid in the limit of
small $h$ and, generally speaking, the problem should be solved
numerically. Below, we consider this problem in the framework of
the DNLS model with a single on-site nonlinear defect.

The paper is organized as follows. In Sec. II we introduce our
model and discuss the analytical results obtained for the family
of the stationary states localized near the defect site. Then, in
Sec. III we study numerically symmetric localized modes and their
symmetry-breaking instability, including the analysis of the
linear eigenvalue problem. Section IV includes the results of the
instability-induced dynamics of localized modes and the study of
the families of asymmetric localized modes which appear as stable
states due to a balance of the defect repulsion and the effective
trapping potential of the lattice. We also find the families of
asymmetric modes and demonstrate their stability. Finally, Sec. V
concludes the paper.

\section{Model and Analytical Results}

We consider a DNLS model with a nonlinear impurity on one site of
the lattice. To examine the interplay between discreteness,
nonlinearity and disorder, we study the case of a focusing
nonlinearity in the waveguide array with a repulsive nonlinear
defect. The DNLS equation describing our model can be written in
the following dimensionless form,
\begin{eqnarray}
i \dot{u}_n + \frac{C}{2}\Delta_2 u_n + |u_n|^2 u_n = \alpha
\delta_{n,n_0} |u_n|^4 u_n,
\label{ieq1}
\end{eqnarray}
where $\Delta_2 u_n \equiv (u_{n+1}+u_{n-1}-2 u_n)$, $u_n$
describes the complex envelope of the electric field in the $n$-th 
waveguide, while the dot stands for the spatial
derivative along the array propagation direction (which we treat
here as time $t$).

In this study, we select the quintic nonlinearity for the defect,
in order to differentiate its effect from the lattice as well as
to model non-Kerr nonlinear response often observed in experiment
(see, e.g. Ref. \cite{Mor_et02}). However, we expect that the
results will be qualitatively valid for any type of the power-law
nonlinearity of the order $n$. The generalized nonlinearity of the
defect can also appear in the theory of photonic circuits
\cite{minga}.

The coupling coefficient $C$ can be expressed through the lattice
spacing $h$ as follows $C=1/h^2$, and it characterizes the
effective diffraction in the array. We should notice that the
lattice spacing $h$ may not coincide with the relative distance
between the waveguide in an array, and these values are connected
in a more complicated way \cite{gera,suk_kiv}.

Equation (\ref{ieq1}) stems from the Hamiltonian
\begin{eqnarray}
H= \sum_n \frac{1}{2} \left(  C |u_{n+1}-u_n|^2 - |u_n|^4 \right) +
\frac{\alpha}{3} |u_{n_0}|^6,
\label{ieq2}
\end{eqnarray}
upon setting its derivative with respect to $u_n^{\star}$ equal to
$i \dot{u}_n$.

The linear waves of the form $u_n=u_0 \exp(i \Omega t - iknh)$
have the (phonon) spectrum $\Omega = -(2/h^2) \sin^2(kh/2)$
consisting of a band of negative frequencies ($\Omega <0$).

In the homogeneous (i.e., no impurity or equivalently $\alpha=0$),
continuum (i.e., $h \rightarrow 0$) limit, the model (\ref{ieq1})
has an exact localized solution of the form (in the long-wave
approximation,  i.e. $h\eta \ll 1$)
\begin{eqnarray}
u (x,t)= \frac{\eta \; e^{i \Lambda t}} {\cosh[\eta (x-\xi)]},
\label{ieq3}
\end{eqnarray}
where $\Lambda=\eta^2/2$, $\Lambda$ is the frequency of the
breather (and should be positive to avoid resonances with the
linear spectrum of plane waves), and $\eta$ essentially defines an
effective mass of the breather excitation in the long wave
approximation according to $M_{eff} = 2 \eta$.

A natural approach to the study of a discrete model is to use the
expression (\ref{ieq3}) as an {\it ansatz solution} in the
Hamiltonian (\ref{ieq2}) in order to obtain information about the
static solution of Eq. (\ref{ieq1}) and its stability.  The last
term of Eq. (\ref{ieq2}) is then easy to evaluate in the framework
of the {\it ansatz} (\ref{ieq3}). On the other hand, the first two
terms (in the parenthesis) of Eq. (\ref{ieq2}) can be evaluated
using the Poisson summation formula \cite{br10} which states that
\begin{eqnarray}
\sum_{n=-\infty}^{\infty} f(\beta n)=\frac{\sqrt{2 \pi}}{\beta}
\sum_{m=-\infty}^{\infty} F \left(\frac{2 m \pi}{\beta} \right),
\label{ieq4}
\end{eqnarray}
where $F$ is the Fourier transform of $f$,
\begin{eqnarray}
F(k)=\frac{1}{\sqrt{2 \pi}} \int_{-\infty}^{\infty} f(x) e^{i k x}
dx. \label{ieq5a}
\end{eqnarray}

Using the above remarks and formulas for the Hamiltonian
(\ref{ieq2}), we can evaluate $H=H(\xi;\{\eta,h\})$ (up to
constant, $\xi$-independent  terms) as the following
\begin{equation}
\label{ieq5}
 H= H_p + H_{\rm int},
\end{equation}
where
\[ H_{\rm p} =- \frac{4 \pi^2}{h^2} \sum_{m=1}^{\infty}  \frac{ m\; \cos(2 m \pi
\xi/h)}{\sinh(m \pi^2/\eta h)} \left[-\frac{1}{h^2}
+\frac{\eta^2}{3}\left(1+\frac{m^2 \pi^2}{\eta^2 h^2}\right)
\right],
\]
\[ H_{\rm int} = \frac{\alpha \eta^6}{3 \, \cosh^6(\eta \xi)}.
\]
Notice that the first part of the expression pertaining to the
summation assumes the correct asymptotic (i.e., small $h$) form
used e.g., in Eq. (3.5) in \cite{KKM}, as well as in references
therein. Finally, it is worth mentioning that the series in Eq.
(\ref{ieq5}) has terms which are exponentially smaller with
respect to the leading order terms and hence even keeping a small
number of terms in the series should yield reasonably accurate
results. In the leading order approximation, the Hamiltonian can
be approximated as follows
\begin{eqnarray}
H \simeq -\frac{16{\pi}^2}{h^4} e^{-{\pi}^2 / h\eta} \cos(2 \pi
\xi /h) + \frac{\alpha {\eta}^6}{3 \cosh^6 \eta \xi} \label{knew1}
\end{eqnarray}

The criterion for obtaining a steady state in the reduced variable
formalism of $H=H(\xi;\{\eta,h\})$ requires that
\begin{eqnarray}
\frac{\partial H}{\partial \xi}|_{\xi=0} =0. \label{ieq6}
\end{eqnarray}
It is worth noting that the fact that the solutions of Eq.
(\ref{ieq6}) will be static solutions of Eq. (\ref{ieq1}) has been
{\it rigorously} proved recently by Kapitula \cite{Kapitula},
through the use of a Lyapunov-Schmidt reduction. Given that the
{\it ansatz} of Eq. (\ref{ieq3}) is not exact, this result should
be interpreted in the orbital sense (i.e., there are orbits in the
configuration space which are {\it close} to the approximate
solution of Eq. (\ref{ieq6}) in conjunction with Eq. (\ref{ieq3}),
in the appropriate norm). In particular, in this case one can
directly observe that $\xi=0$ will always be a solution of Eq.
(\ref{ieq6}). Additional solutions are present close to $\xi=n h$
($n \in {\cal Z}$), and unstable ones close to $\xi=(n+1/2) h$. In
the homogeneous limit, the pulse-like steady states are exactly at
these points, as is well-known, but now the presence of the
impurity modifies their exact location, albeit slightly due to the
exponential decay of the impurity potential away from its location
at  $\xi=0$.

Of particular interest is the stability of the solution at
$\xi=0$. One naturally expects that when the impurity is
sufficiently strong and repulsive, it will {\it reverse} the local
stability picture close to $\xi=0$, converting the relevant local
energy minimum into a local energy maximum. We should note here
that as per the positive definite nature of the Hessian matrix
 in the focusing case (for a
single pulse), the local minima of the energy landscape
$H=H(\xi;\{\eta,h\})$ will correspond to stable solutions. The
opposite will be true for local maxima.
The Hessian matrix here is given by
$\partial P/\partial \Lambda$, where $P$ is the power or (squared) $L^2$
norm of the solution $P=||u||_2^2$ \cite{Peli}.
Recall that the $L^p$ norm is defined as
$||u||_p=( \sum_{n} |u_n|^p)^{1/p}$. The $L^{\infty}$ norm (used also below)
is $||u||_{\infty}=\max_n |u_n|$. $P$ can be thought of as
the number of elementary excitations present in the field.
In the optics literature, this is often called the
``energy'' of the solution.

Hence, the instability
criterion in the reduced, one degree of freedom dynamical system
for the relevant steady state will be
\begin{eqnarray}
\frac{\partial^2 H}{\partial \xi^2}|_{\xi=0}<0,
\label{ieq7}
\end{eqnarray}
which can be restated in the form of the asymptotic approximation,
through the leading order term of the series in Eq. (\ref{ieq5}),
as follows $\alpha > \alpha_c$, where
\begin{eqnarray}
\alpha_c = \frac{32 \pi^4}{h^6 \eta^8} e^{-\frac{\pi^2}{\eta h}}. \label{ieq8}
\end{eqnarray}
In terms of the full series, the critical value reads
\begin{eqnarray}
\alpha_c = \frac{8 \pi^4}{\eta^8 h^6} \sum_{m=1}^{\infty}
\frac{m^3}{ \sinh(m \pi^2/\eta h)} \left[
-\frac{1}{h^2}+\frac{\eta^2}{3}\left(1+\frac{m^2 \pi^2}{\eta^2
h^2}\right) \right], \label{ieq9}
\end{eqnarray}
which can be easily confirmed to have as leading order behavior Eq.
(\ref{ieq8}).

The result (\ref{ieq9}) describes the stability threshold of the
discrete mode localized at the impurity site $n=0$, so that when
the impurity is sufficiently strong and repulsive, it will reverse
the local stability of the mode close to $\xi=0$, converting the
relevant local energy minimum into a local energy maximum.
However, the results (\ref{ieq6}), (\ref{ieq7}) are more general,
and they suggest the existence of many different stable stationary
states located at certain distances from the impurity according to
the local minima of the effective energy, the minima are produced
by the effective periodic potential due to the model discreteness.
For all such states, the physical picture based on the reduced
Hamiltonian is valid only approximately, and more detailed
numerical studies are required.

\section{Localized modes and their stability}

We now turn to numerical results.
We numerically constructed solutions of Eq.(1), for the chain with
$N=200$ sites,
localized at the impurity site $n_0=100$,  using a Newton method with an
initial condition given by Eq. (\ref{ieq3}), assuming $\Lambda=0.5$
($\eta=1$).
Upon convergence, linear
stability analysis of the resulting pulses was performed.
This was done by linearizing using
 the ansatz: $u_n=u_{sol} + \epsilon \exp(i\Lambda t)
[a_n \exp(-i\omega t)+ b_n \exp(i\omega^*t)]$;  $u_{sol}$
is the exact solution (around which we are linearizing) and
the $\omega$'s denote the linearization eigenfrequencies.
Then to O($\epsilon$) we obtain the
relevant (linearization) eigenvalue problem
\[
\omega
\left( \begin{array}{c}
a_{k} \\
b_{k}^{\star} \\
\end{array} \right)
= {\bf J} \cdot
\left( \begin{array}{c}
a_{k} \\
b_{k}^{\star} \\
\end{array} \right), \\
\]
where ${\bf J}$ is the linear stability (Jacobian) matrix of the
form
\[
{\bf J}=
\left( \begin{array}{cc}
\frac{\partial {F_i}}{\partial u_j} & \frac{\partial {F_i}}{\partial u_j^{\star}} \\
-\frac{\partial {F_i^{\star}}}{\partial u_j} & - \frac{\partial {F_i^{\star}}}{\partial u_j^{\star}} \\
\end{array} \right), \\
\]
where $F_i=-\frac{1}{2}C (u_{i+1}+u_{i-1}-2 u_i) - u_i^2
u_i^{\star} + \alpha u_i^3 {u_i^{\star}}^2 \delta_{i,n_0}$. When
the corresponding eigenfrequencies $\omega$ are real, then the
solution is linearly stable. On the contrary, the presence of a
mode with a nonzero imaginary part in its eigenfrequency denotes
the presence of an instability.

\begin{figure}[tbp]
\epsfxsize=9.0cm \centerline{\epsffile{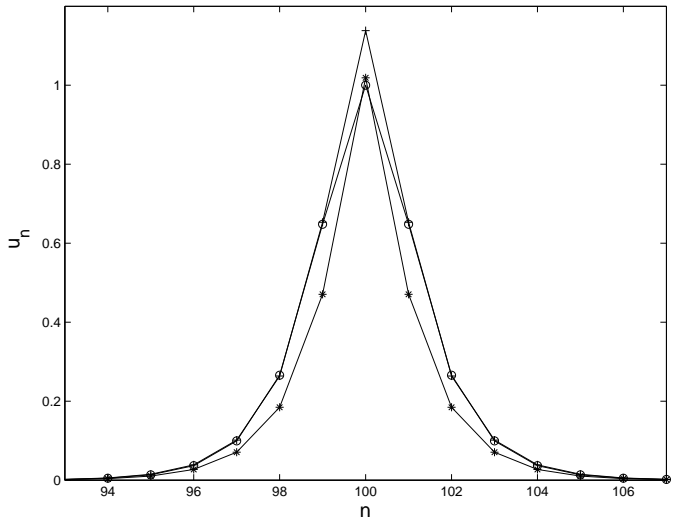}}
\caption{Nonlinear localized mode in the model (\ref{ieq1}) for
$h=1$. Solid line with circles depicts the continuum solution
(\ref{ieq3}). Stars and pluses denote the discrete solution
($h=1$) at $\alpha =0$ and $\alpha =\alpha_c  \approx 0.22$,
respectively. In the discrete case, the solid line is a guide for
eye only. } \label{ifig1}
\end{figure}

Figure \ref{ifig1} shows a typical mode profile obtained in
different cases. These soliton profiles are similar to those
discussed in the long-wave approximation in Refs.
\cite{bogdan,sukh,kos_kov}. For $\alpha=0$ (homogeneous case), the
circles denote the discretization of the continuum profile in
accordance with Eq. (\ref{ieq3}). The corresponding {\it exact}
discrete homogeneous solution is shown by the stars. These results
are shown for the lattice spacing parameter $h=1$. Notice, that
discreteness induces a narrowing and increase of amplitude with
respect to the continuum pulse. Finally, the pluses indicate the
solution at the onset of the instability (that occurs as $\alpha$
is increased), for $\alpha=0.22$. An interesting observation is
that at the onset of the instability, the pulse has essentially
regained its continuum shape {\it but} for a significantly larger
amplitude at the central site.

\begin{figure}[tbp]
\epsfxsize=9.5cm \centerline{\epsffile{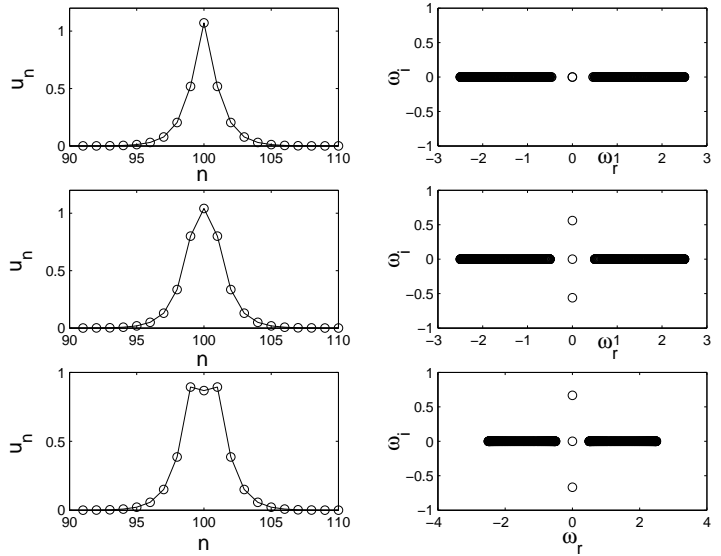}}
\caption{Different localized modes (left panels) and their
stability properties (right panels) in the model (\ref{ieq1}) for
$h=1$. Results of the linear stability analysis are given in the
form of the spectral plane $(\omega_r,\omega_i)$ showing the real
and imaginary part of the eigenfrequency, respectively. The top
row is for $\alpha=0$, the middle is for $\alpha=0.3>\alpha_c$,
while the bottom is for $\alpha=0.5$.} \label{ifig2}
\end{figure}

\begin{figure}[tbp]
\epsfxsize=9.5cm \centerline{\epsffile{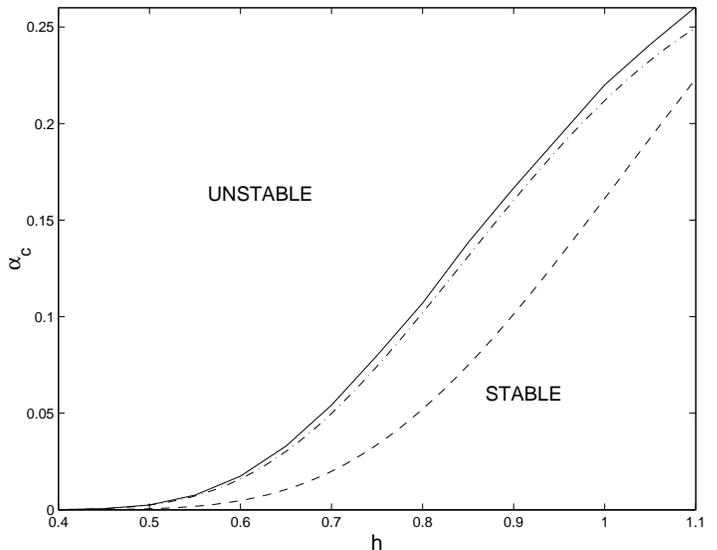}}
\caption{Threshold of the instability onset for the nonlinear
impurity mode centered at the impurity site, shown as the critical
impurity strength $\alpha_c$ vs. the lattice spacing $h$. Solid
line--the full numerical result, dashed line is given by Eq.
(\ref{ieq8}), the dash-dotted is given by Eq. (\ref{ieq9}).}
\label{ifig3}
\end{figure}

A more detailed study of the mode instability that occurs as
$\alpha$ is increased is presented in Figs. \ref{ifig2} and
\ref{ifig3}. Figure \ref{ifig2} shows three different examples of
the discrete localized mode and the corresponding linear stability
results. The top panel is for $\alpha=0.1$, in the subcritical
case, where the stability of the mode can be determined by the
absence of imaginary eigenfrequencies (see the top right panel in
Fig. \ref{ifig2}). The middle row shows a slightly supercritical
value of the impurity strength ($\alpha=0.3$, while $\alpha_c
\approx 0.22$), while a strongly unstable, supercritical mode for
$\alpha=0.5$ is shown in the bottom panel. It is important to
remark here that a perhaps counter-intuitive result of our
numerical investigations is that the instability point is {\it
not} coincident with the point (in parameter space) where the
symmetric mode becomes two-humped. We have generically observed
that the mode becomes two-humped for considerably larger (than the
instability threshold) values of the defect strength $\alpha$.

Similar calculations were performed for different values of the
discreteness parameter $h$ and a two-parameter stability diagram
is shown in Fig. \ref{ifig3}. Above the different curves the
(centered at $\xi=0$) solutions are unstable, while the opposite is
true below the curves. The solid line denotes the exact numerical
result, the dashed line is the prediction of Eq. (\ref{ieq8}),
while the dash-dotted line is the prediction of Eq. (\ref{ieq9})
including $m=50$ terms in the summation. One can observe that
even though the asymptotic result gradually fails (as it should)
for larger $h$, the full series prediction of Eq. (\ref{ieq9})
remains very close to the fully numerical result. The difference
can be well accounted for by the approximations involved in the
continuum {\it ansatz}.

\begin{figure}[tbp]
\epsfxsize=9.5cm \centerline{\epsffile{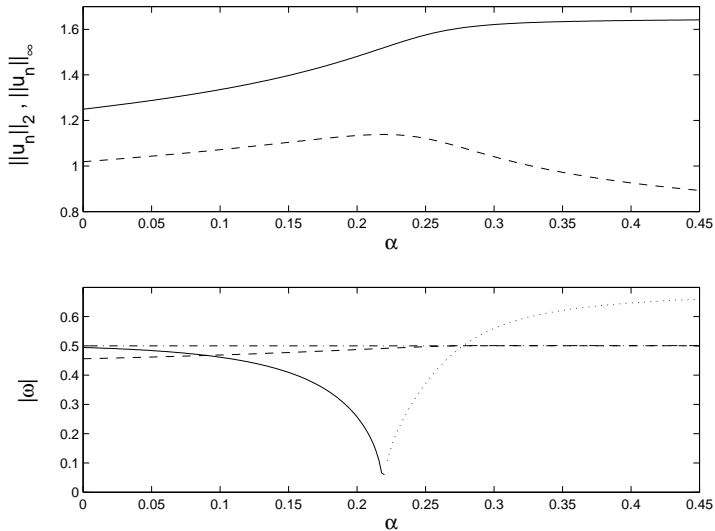}}
\caption{Top: the
$L^2$ (solid line) and $L^{\infty}$ (dashed line)
norms of the localized mode vs. $\alpha$.
Bottom: the relevant point spectrum eigenvalues of the problem
continued over $\alpha$. The solid line shows the pinning (or
translational) mode that becomes unstable (dotted line) after
$\alpha_c \approx 0.22$. The dashed line shows the breathing (or edge)
mode, while the dash-dotted line indicates the band edge of the
continuous spectrum.} \label{ifig4}
\end{figure}

Additional insight on the appearance of the instability is given
in Fig. \ref{ifig4}, where we examine the mode stability for
different values of $\alpha$ (at $h=1$). The bottom panel of the
figure shows the trajectory of the main eigenmodes of the point
spectrum of the linearization around the solution. The solid line
shows the translational or pinning (antisymmetric) mode, while the
dashed line shows the edge or breathing (spatially symmetric) mode
\cite{joh}. Notice that for $h=1$ (cf. Fig. 1 of Ref. \cite{joh}),
the pinning mode has a larger frequency than the breathing mode.
The band edge of the continuous spectrum which lies at
$\Lambda=0.5$ \cite{joh} is shown by a dash-dotted line. As
$\alpha$ is increased, it is clear that the breathing mode is
``repelled'' by the presence of the impurity towards the phonon
band edge. On the contrary, the antisymmetric mode gradually
approaches the origin and becomes unstable (dotted part of the
relevant curve) for $\alpha>0.22$ (thereafter, for
$\alpha>\alpha_c$ there appears an imaginary pair of
eigenfrequencies). Perhaps it is even more interesting to examine
the occurrence of this instability in the light of the top panels
of the figure. The latter show the $L^2$ and
$L^{\infty}$ norms of the solution as a function of $\alpha$.
It
is noteworthy that the occurrence of the instability coincides
with the change of concavity of the $L^2$ norm, while it also
coincides with a maximum of the $L^{\infty}$ norm. The above
results suggest two additional numerical criteria that can be used
to identify  the appearance of the instability, namely
\begin{eqnarray}
\frac{\partial^2 ||u||_2}{\partial \alpha^2}= 0,
\label{ieq10}
\end{eqnarray}
and
\begin{eqnarray}
\frac{\partial ||u||_{\infty}}{\partial \alpha}=0; \;\;\;
\frac{\partial^2 ||u||_{\infty}}{\partial \alpha^2}<0.
\label{ieq11}
\end{eqnarray}

\section{Dynamical effects and asymmetric modes}

\begin{figure}[tbp]
\epsfxsize=9.5cm \centerline{\epsffile{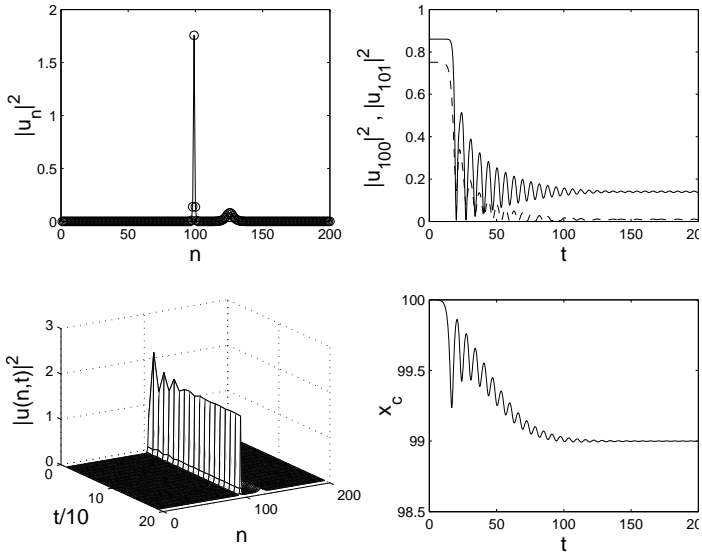}}
\caption{Development of the mode instability for $\alpha=0.3$ and
$h=1$. The top left panel shows the mode spatial profile at
$t=200$. The top right shows the modulus squared of the central
site and of the site nearest to it, indicating relaxation to an
exponentially localized configuration centered around the
neighboring minimum (at $n \approx 99$). The bottom left panel
shows a space time plot of the evolution of the modulus squared of
the field, while the bottom right indicates the relaxation of the
localized mode center from site $x_c \approx 100$ to the one
with $x_c \approx 99$.}
\label{ifig5}
\end{figure}

To examine the dynamical development of the instability, once the
solution becomes unstable for the supercritical values of $\alpha
> \alpha_c$, we have performed direct numerical simulations of Eq.
(\ref{ieq1}) with an initial condition consisting of an {\it
exact, unstable} localized mode for $\alpha=0.3$, perturbed by
small noise (of uniform distribution, and amplitude ranging from
0.0001 to 0.1). The scenario that is described was found to be
generic for such perturbations of the unstable solution. It was
thus found that as is shown in the top left panel of Fig.
\ref{ifig5} for $t=200$, the unstable mode sustains a symmetry
breaking, which cleaves it into {\em another stable localized
mode} centered essentially at the neighboring (previous)
lattice site (
see also the bottom right panel), and to a
small-amplitude mobile pulse that propagates along the chain and
carries an energy excess. It is worth noting that the original
(unstable) eigenmode had 3 ``main'' sites oscillating at an amplitude
of $\approx 0.9$, while the resulting breather is very strongly localized
at a single site (of amplitude $\approx 1.7$).

\begin{figure}[tbp]
\epsfxsize=9.5cm \centerline{\epsffile{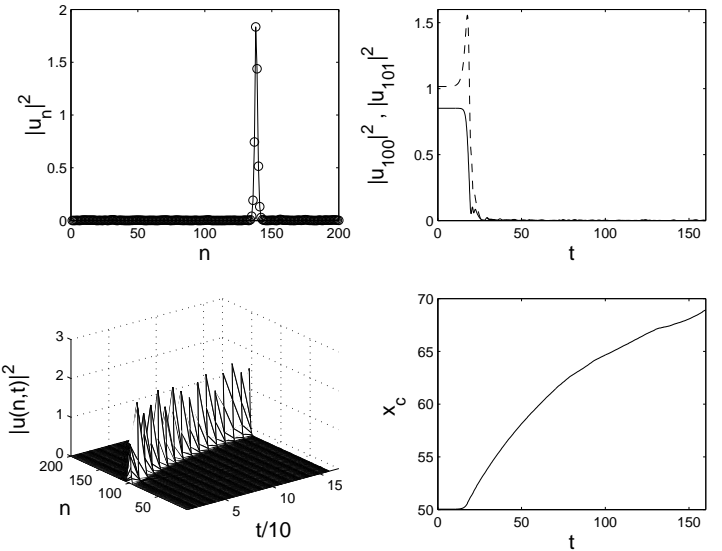}}
\caption{Same as
in Fig. \ref{ifig5} but for $\alpha=1$ and $h=0.5$. Instability
initiates the motion of a localized mode along the chain.}
\label{ifig6}
\end{figure}

The trapping of the unstable mode by the neighboring site is
consistent with the physical picture described by the effective
Hamiltonian (\ref{ieq5}), when the mode overcomes the neighboring
Peierls-Nabarro barrier and becomes localized by the next minimum.
However, for smaller values of the lattice spacing $h$, when the
Peierls-Nabarro barrier becomes negligible, the unstable localized
mode can start its motion through the lattice under the action of
the initial perturbation and as a result of the development of the
symmetry-breaking instability. An example of the latter type is
shown in Fig. \ref{ifig6} for the defect strength $\alpha=1$ and
$h=0.5$.

\begin{figure}[tbp]
\epsfxsize=9.5cm \centerline{\epsffile{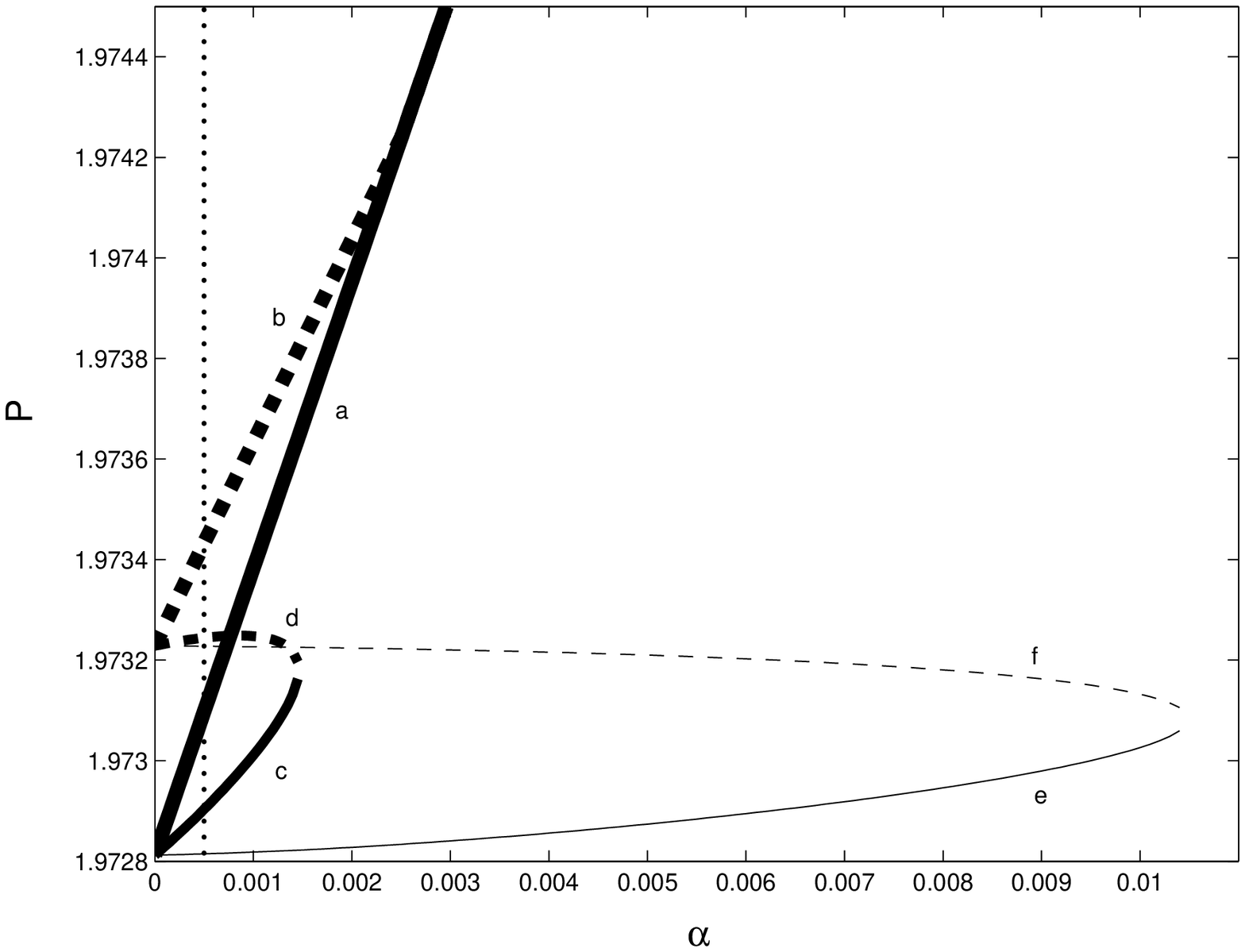}}
\caption{Bifurcation diagram of the asymmetric localized impurity
modes. The thickest solid line (branch a) corresponds to the mode
centered at $n_0=100$, while the corresponding dashed line (branch
b), to the mode centered at $n_0=99.5$. The medium thickness solid line and
the corresponding dashed one (branches c and d) are centered, for
$\alpha=0$ at $n_0=99$ and $n_0=98.5$ respectively. The
corresponding saddle-node bifurcation resulting at the
disappearance of these branches occurs for $\alpha \approx
0.00147$. Finally, the thin solid and dashed lines (branches e and
f) correspond to the cases of $n_0=98$ and $n_0=97.5$,
respectively. The saddle-node bifurcation occurs here at $\alpha
\approx 0.0105$. The dotted line at $\alpha=0.0005$ indicates the
defect strength for which the mode profiles of the different
branches are shown in Fig. \ref{ifig8}. In the
homogeneous case ($\alpha=0$) the array
of minima and maxima of the Peierls barrier occurs for $x_c=m$ and
$x_c=m+1/2$ ($m \in {\cal Z}$). Hence, the points with $x_c=100, 99, 98,
97 \dots$ have the same norm
(the same is true for the branches with $x_c=99.5, 98.5, 97.5 \dots$)
for $\alpha=0$. In the presence of the impurity (e.g., for $\alpha \neq 0$),
this shift invariance is broken. Thus we use the ``original''
(e.g., for $\alpha =0$) center position of the branch to classify them
when $\alpha \neq 0$.}
\label{ifig7}
\end{figure}

One of the most interesting findings of our study of the time
evolution as a  result of the mode instability is the effective
``relaxation'' of the unstable mode towards the mode centered at
the neighboring (to the defect) site. The latter result prompted
us to explore the modes in the vicinity of the defect. Due to
symmetry, our study was restricted to modes on one side of the
impurity site. The bifurcation diagram of the relevant branches of
solutions is shown in Fig. \ref{ifig7} obtained for $h=0.5$. We
have observed the relevant phenomenology to be generic, but report
it for this value of the lattice spacing that necessitates
narrower ranges of parameter sweeps (i.e., the same observations
can be obtained for larger $h$, but due to the weaker coupling,
they occur for considerably larger values of the defect strength
$\alpha$).

\begin{figure}[tbp]
\epsfxsize=9.5cm \centerline{\epsffile{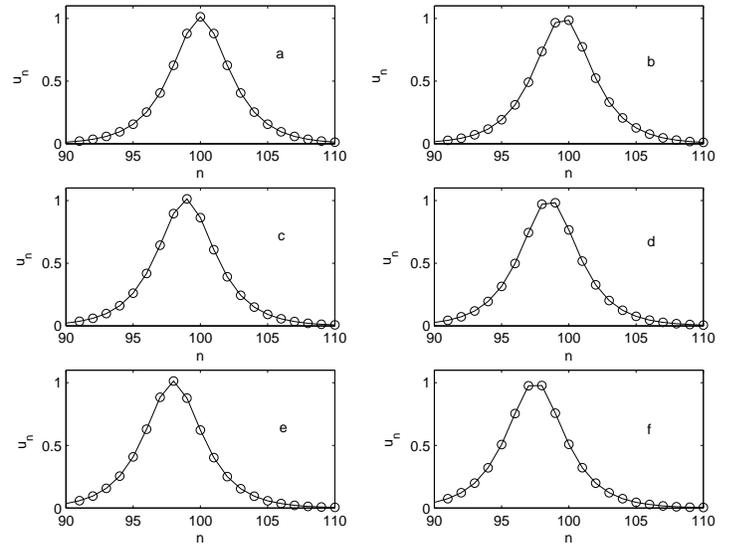}} \caption{Spatial
profiles of the localized modes corresponding to the branches (a)
to (f) in Fig. \ref{ifig8} for the defect strength
$\alpha=0.0005$. A repulsive impurity is located at the site
$n_0=100$.} \label{ifig8}
\end{figure}

The natural starting point for our analysis is the continuation
from the case of an homogeneous lattice ($\alpha=0$), where the
stationary modes are well-known to exist on a lattice site and
between two consecutive lattice sites. Hence, in the homogeneous
case we consider the branches centered at $n_0=100$ (branch a),
$n_0=99.5$ (branch b), $n_0=99$ (c), $n_0=98.5$ (d), $n_0=97.5$
(e) and $n_0=97$ (f). We observe that {\it at} the instability
point $a_c \approx 0.00252$, the stable branch a and the unstable
branch b {\it merge}. The resulting branch is always unstable
thereafter. For any other branch apart from the central one, we
have found that the relevant stable (node) and unstable (saddle)
solutions exist for a parameter interval that depends on the
distance of the central site of the branch from the defect site;
naturally, the branches whose central site is more remote from the
defect are less affected by it (and exist for larger intervals of
defect strengths). These branches eventually disappear in
saddle-node bifurcations, as is observed for branches c and d and
e and f in Fig. \ref{ifig7}. A set of spatial profiles
corresponding to the different branches are shown in Fig.
\ref{ifig8} for the different branches of Fig. \ref{ifig7} and for
$\alpha=0.0005$ (corresponding to the vertical dotted line in Fig.
\ref{ifig7}).

Disappearance of the relevant branches of static solutions is
consonant with the mobility of the unstable modes for smaller
values of $h$ observed in Fig. \ref{ifig6} above. For smaller
values of $h$, the neighboring branches disappear for smaller
defect strengths, hence allowing for no static solutions in the
vicinity (in configuration space) of the unstable solution,
thereby resulting in the mobility of the latter (i.e., the
solution cannot be trapped in neighboring potential wells as these
have disappeared). However, in the opposite case of large $h$, the
lattice is strongly discrete and an adiabatic growth of the defect
strength $\alpha$ leads to the mode motion from one stable state
to its neighboring state (via an intermediate unstable mode)
through a cascade of the bifurcation points.

\section{Conclusions}

We have studied the existence and stability of nonlinear localized
modes in the framework of the DNLS model. We have used a
variational formalism to obtain the analytical result for the
occurrence of an instability due to the interplay between
attractive nonlinearity and repulsive impurity. We have examined
the variational prediction by means of numerical bifurcation
theory, linear stability analysis, as well as direct integration
of the discrete lattice model.  We have found the variational
prediction (and its refinements) to be in reasonable agreement
with the numerical results, attributing the discrepancies to the
variation of the original profile with respect to the pre-selected
ansatz. We have observed that the instability develops much before
the solution becomes two-humped, and we have developed numerically
motivated criteria for the identification of the instability. The
study of the dynamical evolution of the unstable modes revealed
the potential for mobile localized modes for smaller values of the
lattice spacing, while it results in relaxation to the neighboring
potential wells and the corresponding stable, static solutions.
Finally, the bifurcation diagram has been explored as a function
of the defect strength, using continuation from the homogeneous
lattice case. We have found that for the branch corresponding to
the defect site, the stable and nearest unstable solution branches
merge at the instability point (producing a thereafter always
unstable branch), while the neighboring sites corresponding stable
and unstable branches disappear through collision via saddle-node
bifurcations. Such a bifurcation scenario is in a perfect agreement
with the concept of the Peierls potential that affects the
mode motion in discrete systems.

PGK would like to thank J. Cuevas for stimulating discussions and
for bringing to his attention the results of \cite{archilla} prior
to publication. He would also like to acknowledge partial support
through a Faculty Research Grant of the University of
Massachusetts and the Clay Foundation for support through a
Special Project Prize Fellowship and the National Science Foundation
for support through DMS-0204585. YSK acknowledges a support from
the US Air Force-Far East Office and the Australian Research
Council. ASK would like to
acknowledge partial support from Royal Swedish Academy of Science.

\end{multicols}
\end{document}